# Tunable nonlinear damping in parametric regime

*Parmeshwar Prasad\*, Nishta Arora, and A. K. Naik\**

*Centre for Nano Science and Engineering,*

*Indian Institute of Science, Bangalore, India, 560012*

*Nonlinear damping plays a significant role in several area of physics and it is becoming increasingly important to understand its underlying mechanism. However, microscopic origin of nonlinear damping is still a debatable topic. Here, we probe and report nonlinear damping in a highly tunable $MoS_2$ nano mechanical drum resonator using electrical homodyne actuation and detection technique. In our experiment, we achieve 2:1 internal resonance by tuning resonance frequency and observe enhanced non-linear damping. We probe the effect of non-linear damping by characterizing parametric gain. Geometry and tunability of the device allow us to reduce the effect of other prominent Duffing non-linearity to probe the non-linear damping effectively. The enhanced non-linear damping in the vicinity of internal resonance is also observed in direct drive, supporting possible origin of non-linear damping. Our experiment demonstrates, a highly tunable 2D material based nanoresonator offers an excellent platform to study the nonlinear physics and exploit nonlinear damping in parametric regime.*

Dynamical systems in nature show rich physics demonstrating energy exchange. The systems find its equilibrium mediated by energy exchange. There are several ways through which a system accomplishes energy exchange with its surroundings. The energy exchange could be dissipative in nature and can be modelled as linear damping in the simplest case. However, there are system which shows diverse nonlinear dissipative phenomena[1–3]. For example, nano mechanical systems (NEMS) offer such a system where vibration amplitude can be tuned to an extent to witness nonlinear dissipation[2–5]. In general, vibration amplitude comparable to the length scale of its thickness drives the system into nonlinear regime. In this regime, the dynamical systems offer rich physics dealing with nonlinear dynamics. As the nonlinearity in the system increases the presence of dissipative mechanics play an important role in dynamics of the nano-mechanical systems.

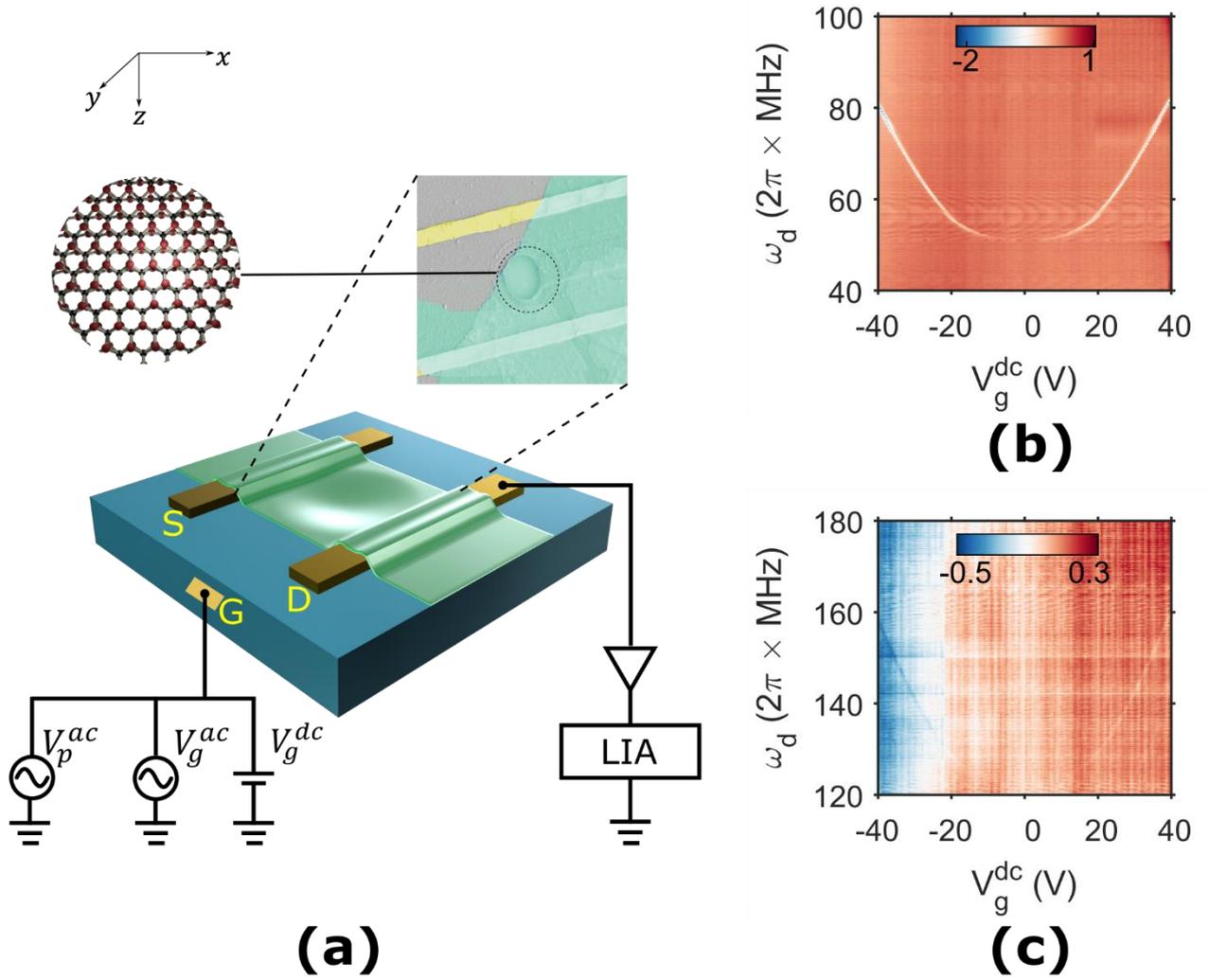

*Figure (1) Characterisation of the device: (a) Inset: SEM image of the device, the two thick yellow lines are source and drain electrodes, the suspended region is 3 μm in diameter. The gate is 300 nm below the suspended membrane. Schematic of the measurement setup: Homodyne capacitive actuation and detection technique. Membrane frequency is tuned using DC gate voltage $V_g^{dc}$, actuated directly using $V_g^{ac}$ at frequency ω and parametrically at 2ω using $V_p^{ac}$. AC voltages are combined using RF power combiner, further DC is combined with the help of a bias-tee. Readout is amplified using low noise amplifier and measured at lock-in amplifier locked at ω. Frequency dispersion with $V_g^{dc}$, two modes $\omega_1$ (b) and $\omega_2$ (c) with frequency tuning >1 MHz/V and $\omega_2 \sim 2\,\omega_1$.*

Nonlinear dissipation has been of interest due to its presence in the application in biology[6], magnetization[7], quantum optics[8] and quantum information technologies[9]. However, source of their origin remains unclear and remains an important area to explore especially in low dimensional nanoresonators. Advancement in the nano-electromechanical system (NEMS) in the last decade has led to exploration of nonlinear dissipation mechanisms [2–4,10–12]. It has been shown that the effect of nonlinear damping increases with vibration amplitude[13]. The nonlinear term can be incorporated in the force equation as $\propto \eta x^2 \dot{z}$ where $\eta$ is the nonlinear damping coefficient, $z$ is the displacement and $\dot{z}$ is the velocity[13]. Though there are several models explaining the microscopic origin of the nonlinear damping, the precise mechanism is not understood [10,14,15]. This is partly due to the challenges involved in isolating and decoupling the mechanism responsible for the nonlinear damping from other nonlinear effects. Nonlinear phenomena are easily achievable in low-dimensional materials due to their extreme sensitivity to the external perturbation. In this context, 2D materials such graphene and transition metal di-chalcogenide (TMDC) provide an excellent platform to probe nonlinear damping. Recently, work by Keskekler et. al[4]. shed light on the possible mechanism of the nonlinear damping through internal resonance (IR) in graphene using the optical actuation and detection method. Though optical method provides a sensitive detection technique, it lacks the tunability of the resonance frequency without heating the device. Heating is one of most common way to dissipate energy.

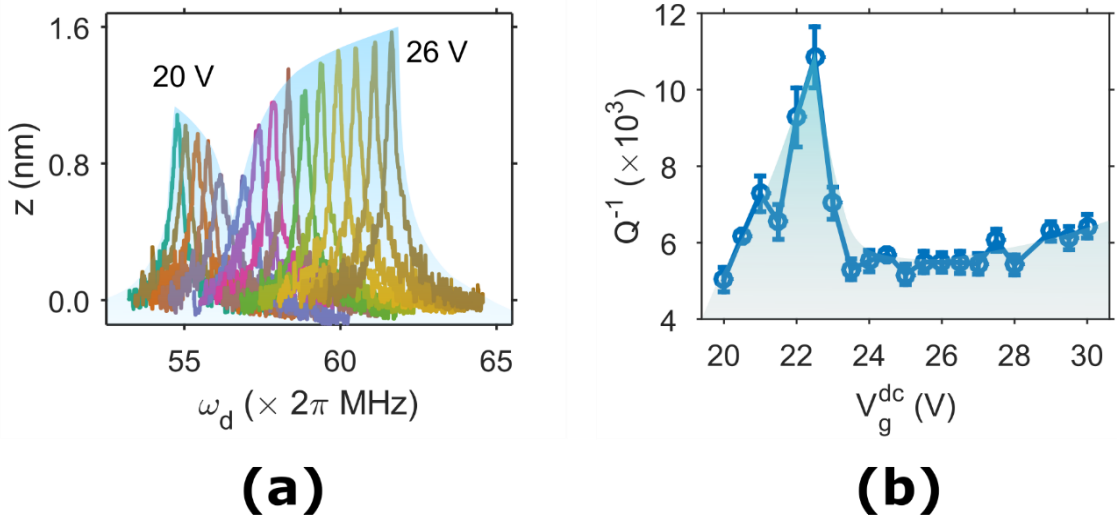

*Figure (2) (a) Amplitude response curve of mode1 ($\omega_1$) at multiple gate voltages. $V_g^{dc} \times V_g^{ac} = 0.2\ V^2$ is constant for the experiments. Shaded region shows the variation of amplitude. Amplitude is minimum at $V_g^{dc} = 22\ V$ (b) Inverse of quality factor at different $V_g^{dc}$, it quantifies the damping in the system. Damping is maximum at $V_g^{dc} = 22\ V$.*

In this work, we demonstrate the nonlinear damping mechanism in a highly tunable molybdenum disulphide ($MoS_2$) device using the electrostatic homodyne capacitive actuation and detection method. The method provides a clean, fast and simple platform to study the nonlinear system[16,17]. Our device shows two prominent modes $\omega_1$ and $\omega_2$ the two modes are $\omega_2 \sim 2\omega_1$. The modes provide favourable scenario to study the nonlinear damping in the parametric regime and probe the dissipation mechanism. In our experiment, we are able to tune the nonlinear damping by changing the applied gate voltage. We find the effect of nonlinear damping considerably increases when the two modes are commensurate with each other in 2:1 ratio. To probe the nonlinear damping, we use parametric pumping technique. Examining the parametric gain, we find that the effect of nonlinear damping reaches maximum in the vicinity of IR. Parametric gain decreases by as low as 83% when the nonlinear damping is maximum. We observe, the enhancement in nonlinear damping is due to strong coupling between the two modes near internal resonance. The experiment shed light into the mechanism of the nonlinear damping into nano mechanical system using a highly tunable NEMS.

Our device is ~ 10 layer thick $MoS_2$ drum resonator suspended over 3 μm diameter trench. The device is strained and actuated by applying voltages ($V_g$) at gate electrode. The gate electrode is 300 nm below the suspended membrane. Figure 1a shows the schematic of the measurement used to measure the vibration of the membrane and for applying parametric pumping. The membrane is strained by applying a DC gate voltage ($V_g^{dc}$) and actuated with an AC voltage ($V_g^{ac}$). The DC and the AC is combined using a bias-tee before applying to the gate electrode. Applied potential difference exerts force on the suspended membrane there by displacing it from the equilibrium position. The displacement of the membrane modulates the geometric capacitance between gate and the membrane, leading to the change in voltage at the drain. The change in drain voltage is amplified using a low-noise amplifier before measuring at the lock-in amplifier. Figure 1b and c show the dispersion of the frequency with applied DC gate voltage. We observe two distinct modes having similar tuning with each other. The two frequencies are $\omega_1 \sim$ 56 MHz and $\omega_2 \sim$111 MHz at around $V_g^{dc} = 23$ V . The modes are interesting due to 2:1 (i.e. $\omega_2 \sim 2\omega_1$) ratio of the frequencies, which enables us to achieve internal resonance by driving the fundamental mode in parametric regime. We can achieve this IR ratio by tuning the gate voltage.

To obtain the elementary characteristics of the device, we drive it with fixed force ($F \propto V_g^{dc} V_g^{ac}$) over a range of gate voltages ($V_g^{dc}$). A constant force allows us to observe and compare the variation of amplitude response curve at different resonant frequency. We drive the resonator with small force such that the response does not show nonlinear behaviour. The linear response can be quantified by the symmetry of the peak about the resonance frequency. In a nonlinear regime a hysteresis in the frequency response curve is observed. Figure 2a shows amplitude response with frequency at different gate voltages. We observe that amplitude decreases with increase in gate voltage and reaches minimum value around $V_g^{dc} \sim 22$ V. In a linear system, a constant force results in a constant amplitude and linewidth due to linear dissipation. Reduction in amplitude for a constant applied force indicates enhanced dissipation. The line width of the response curve is shown in the figure 2b, it shows damping is maximum at 22 V. To probe the enhancement in damping, we perform parametric pumping by tuning stiffness of the membrane. Parametric pumping offers advantage in probing a resonance mode by applying a force away from the resonant frequency and study the coupling of modes. Parametric pump can be used alone or in a combination with direct drive. Parametric pump works in a principle that it provides energy to overcome the linear damping of the system by modulating the stiffness.

We use the following equation of motion to describe the parametric amplification in our system[13]:

$$m\ddot{z} + \gamma \dot{z} + m\omega_0^2 z + \eta \dot{z} z^2 + \alpha_{eff} z^3 = F(\omega\, t) + F_p(2\omega\, t)\, z \qquad (1)$$

Where $m$ is mass of the system $\gamma$ is the linear damping, $\omega_0$ is the resonance frequency, $\eta$ is the nonlinear damping coefficient, $\alpha_{eff}$ is the effective Duffing nonlinearity coefficient associated with geometric nonlinearity. $F$ is the direct force applied at the frequency $\omega$, $F_p$ is the parametric force acting at the system at the frequency $2\omega$ and $z$ is the vibration amplitude.

To study parametric amplification, we apply an alternating voltage ($V_p^{ac}$) at twice the resonance frequency of the fundamental mode at the gate in addition to $V_g^{ac}$ (see Figure 1a). In a parametric system, there exists a critical pumping force beyond which the system overcomes linear damping and reaches instability region also known as Mathew's tongue[13,18]. The critical pumping force can be measured by sweeping $V_p^{ac}$ at $2\omega$. Figure 3 shows amplitude response using parametric drive beyond critical pump voltage ($V_p^{ac0}$) for different $V_g^{dc}$. Beyond the critical pumping, the amplitude response is governed by the nonlinear parameters and offers an excellent regime to probe the nonlinear coefficients. There are two dominant nonlinearities present in our device namely Duffing nonlinearity ($\alpha_{eff}$) and nonlinear damping ($\eta$). We observe that $\alpha_{eff}$ changes from negative to positive, it changes sign around $V_g^{dc} \sim 24$ V. It is interesting for the fact that this allows us to probe the effect of nonlinear damping by minimizing the effect of the prominent geometric nonlinearity. Figure 3d shows the computed value of $\alpha_{eff}$ using continuum model (see supplementary for derivation and computation). The variation of computed $\alpha_{eff}$ matches well with our experimental findings. Figure 3a also shows decreases in amplitude response at around $V_g^{dc} \sim 24$ V. The variation in amplitude response similar to the experimental findings shown in figure2a. In the past, it has been shown that the amplitudes in parametric regime are saturated by nonlinearities present in the system[19]. The current device provides favourable platform to investigate the effect of the nonlinear damping while suppressing the effect of $\alpha_{eff}$. To probe the nonlinear damping, we perform parametric amplification and observe the amplitude gain in the system.

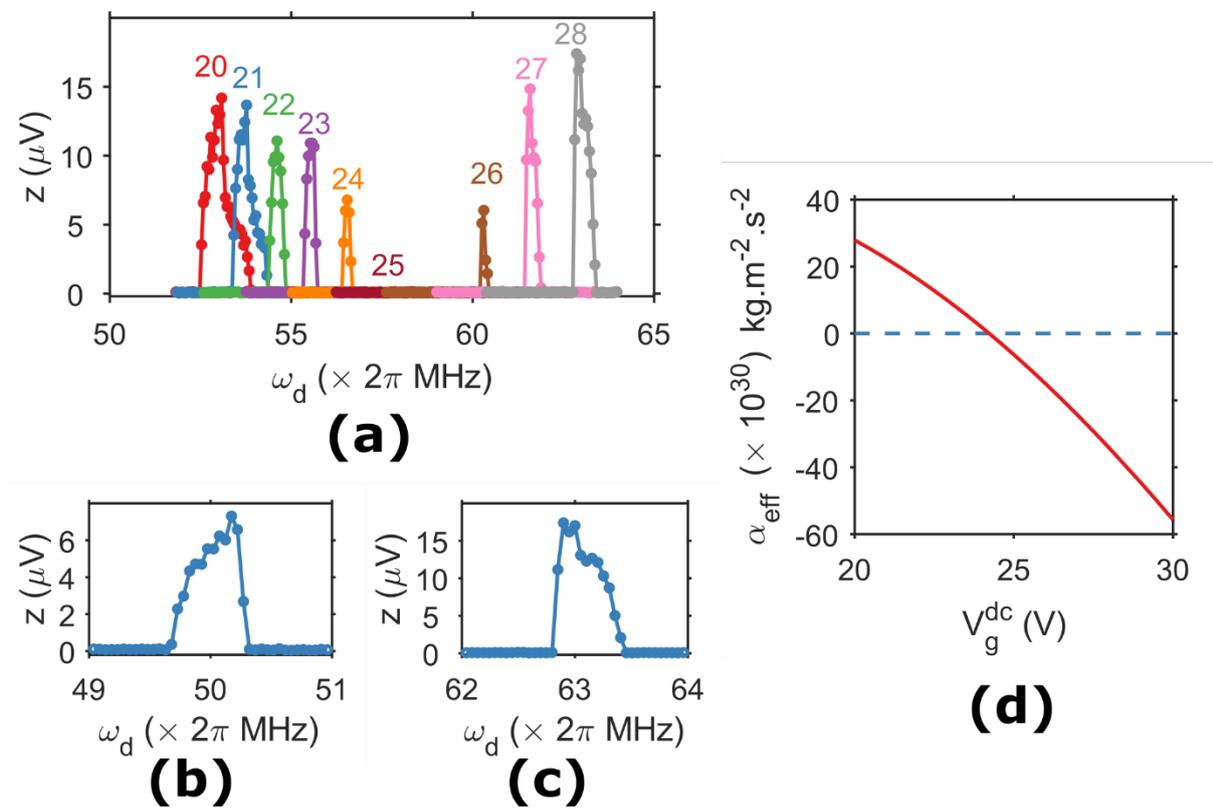

*Figure 3 Tuning of geometric nonlinearity (a) Amplitude response with upward frequency sweep for different $V_g^{dc}$ with $V_p^{ac} = 400\ mV$ and $V_g^{ac} = 0$. Change in the shape of amplitude response curve shows variation in geometric nonlinearity. Amplitude response at (b) $V_g^{dc} = 16$ V and (c) $V_g^{dc} = 28$ V. (d) Blue line is effective Duffing nonlinearity ($\alpha_{eff}$) calculated using continuum model, it crosses the $\alpha_{eff} = 0$ (red line) around $V_g^{dc} = 24$ V, which matches the experimental observation of in (a).*

In parametric amplification, the gain ($G = \frac{z_{on}}{z_{off}}$) is defined as the ratio of amplitude when $V_p^{ac}$ is ON to OFF. Figure 4a shows parametric gain with $V_p^{ac}$ for different $V_g^{dc}$. It shows that gain increases steadily and get saturated after the critical pumping voltage. The maximum gain $G_{max}$ obtained before the critical pumping is shown in the figure 4b. It shows that the $G_{max}$ decreases significantly in the region $V_g^{dc} = 22$ V to $V_g^{dc} = 26$ V. The decrease in the $G_{max}$ is as low as 83% as compared the $G_{max}$ at $V_g^{dc} = 20$ V. To find the nonlinear damping coefficient we use the relation $\eta = \frac{2(F_p Q - 2)}{z_{max}^2}$, where $F_p$ is parametric force, $Q$ quality factor and $z$ is the maximum amplitude[13]. The relation allows us to estimate the coefficient of nonlinear damping independent of any nonlinear parameter in the lowest order. Figure 4e shows the coefficient of the nonlinear damping extracted for different gate voltages.

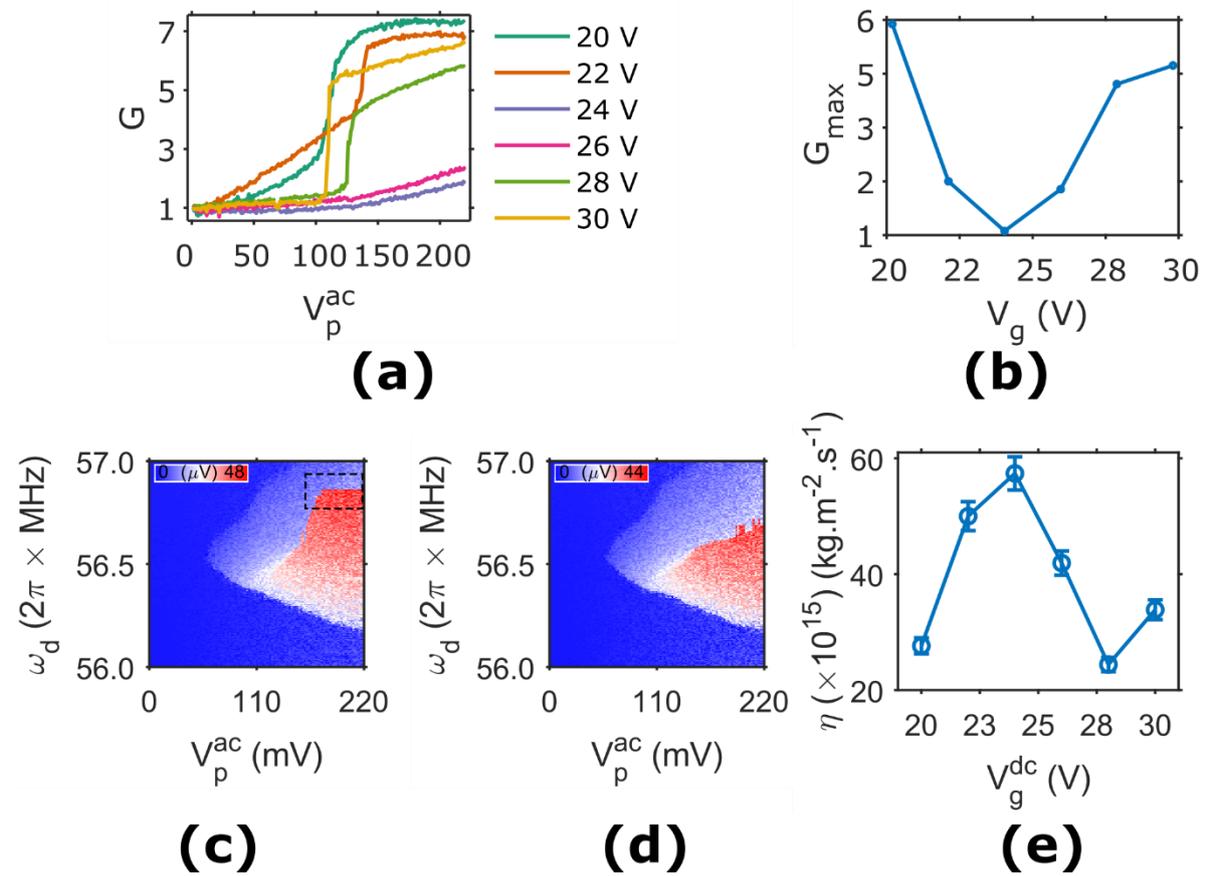

*Figure (4) Parametric amplification: (a) Amplitude gain (G) vs applied pump strength $V_p^{ac}$ for multiple $V_g^{dc}$. Gain increases as pump strength increases, eventually saturating due to nonlinear effects. (b) Maximum gain ($G_{max}$) at the start of instability tongue vs $V_g^{dc}$. $G_{max}$ decreases in the region $V_g^{dc} = 22$ V to $V_g^{dc} = 26$ V. (c)(d) Self oscillation of mode1, response measured at ω is plotted against pump $V_p^{ac}$ applied at frequency 2ω. (c) and (d) are upward*

*and downward frequency sweep respectively. Dashed rectangle shows the region of frequency locking, resulting from 1:2 internal resonance. (e) Coefficient of nonlinear damping η calculated using the parametric model using the equation 1. Higher value of η is observed in the region $V_g^{dc} = 22\ V$ to $V_g^{dc} = 26$ V. It explains the low gain observed in (b) and matches the experimental findings.*

We observe that the nonlinear damping increases significantly around $V_g^{dc} = 24\ V$. Around the gate voltage, we observe the minima of the amplitude and increased damping in case of direct driving (See figure 2a, b). Enhancement of nonlinear damping is evident, as suggested by decrease in amplitude during direct drive and the gain during parametric amplification. Figure 4c,d shows instability region, $V_p^{ac}$ is swept at frequency $2\omega$ and amplitude response is measured at $\omega$. As we increase the parametric pump strength, the direction of peak shifts from lower frequency and to higher frequency, indicating change in the value of $\alpha_{eff}$, which matches our theoretical prediction using continuum model of the drum resonator. Upon further increasing the pump, we observe the frequency locking, we note that this is in the vicinity of favourable 2:1 internal resonance. In the internal resonance, the two modes are coupled to each other, it makes pathway to exchange the energy. In the vicinity of IR, the mode1 dissipates energy to mode2 enhancing the nonlinear damping.

In summary, we study rich nonlinear physics in MoS$_2$ drum resonator in parametric regime. We observe two prominent tunable modes in 50 MHz to 200 MHz range and they are coupled to each other through tension in the membrane. Dominant nonlinear coefficients $\alpha_{eff}$ and $\eta$ are tunable in our device using DC gate voltage. We tune the modes such that it facilitates 2:1 internal resonance and exchange energy with each other. We observe that the nonlinear damping increases significantly in the vicinity of internal resonance, $\eta$ reaching as high as $60 \times 10^{15}$ kg m$^{-2}$s$^{-1}$ in MoS$_2$ drum resonator. In our experiment, minimizing the effect of $\alpha_{eff}$ helps us to probe nonlinear damping efficiently. A highly tunable NEMS device offers versatile control on the nonlinear coefficients and in probing them. Our experiment shows evidence of the microscopic origin of nonlinear damping and shed light in understanding the underlying physics. Parametric regime provides excellent approach to probe the nonlinear physics by tuning the fundamental parameters such as stiffness.

**Acknowledgements:**

We acknowledge funding support from Nano Mission, Department of Science and Technology (DST), India through Grants SR/NM/NS-1157/2015(G) and SR/NMITP-62/2016(G) and from Board of Research in Nuclear Sciences (BRNS), India through Grant 37(3)/14/25/2016-BRNS. P.P. acknowledges scholarship support from CSIR, India. N.A. acknowledges fellowship support under Visvesvaraya Ph.D.Scheme, Ministry of Electronics and Information Technology(MeitY), India. We also acknowledge funding from MHRD, MeitY, and DST Nano Mission for supporting the facilities at CeNSE. We gratefully acknowledge the usage of the National Nanofabrication Facility (NNfC) and the Micro and Nano Characterization Facility (MNCF) at CeNSE, IISc, Bengaluru.

**References:**

[1] A. Eichler, J. Moser, J. Chaste, M. Zdrojek, I. Wilson-Rae, and A. Bachtold, Nat. Nanotechnol. **6**, 339 (2011).

[2] J. Güttinger, A. Noury, P. Weber, A.M. Eriksson, C. Lagoin, J. Moser, C. Eichler, A. Wallraff, A. Isacsson, and A. Bachtold, Nat. Nanotechnol. **12**, 631 (2017).

[3] A. Eichler, J. Chaste, J. Moser, and A. Bachtold, Nano Lett. **11**, 2699 (2011).

[4] A. Keşkekler, O. Shoshani, M. Lee, H.S.J. van der Zant, P.G. Steeneken, and F. Alijani, Nat. Commun. **12**, 1099 (2021).

[5] J. Atalaya, T.W. Kenny, M.L. Roukes, and M.I. Dykman, Phys. Rev. B **94**, 195440 (2016).

[6] M. Amabili, P. Balasubramanian, I. Bozzo, I.D. Breslavsky, G. Ferrari, G. Franchini, F. Giovanniello, and C. Pogue, Phys. Rev. X **10**, 011015 (2020).

[7] B. Divinskiy, S. Urazhdin, S.O. Demokritov, and V.E. Demidov, Nat. Commun. **10**, 5211 (2019).

[8] L. Mandel and E. Wolf, *Optical Coherence and Quantum Optics* (Cambridge University Press, 1995).

[9] Z. Leghtas, S. Touzard, I.M. Pop, A. Kou, B. Vlastakis, A. Petrenko, K.M. Sliwa, A. Narla, S. Shankar, M.J. Hatridge, M. Reagor, L. Frunzio, R.J. Schoelkopf, M. Mirrahimi, and M.H. Devoret, Science **347**, 853 (2015).

[10] S. Zaitsev, O. Shtempluck, E. Buks, and O. Gottlieb, Nonlinear Dyn. **67**, 859 (2012).

[11] V. Singh, O. Shevchuk, Y.M. Blanter, and G.A. Steele, Phys. Rev. B **93**, 245407 (2016).

[12] M. Imboden, O. Williams, and P. Mohanty, Appl. Phys. Lett. **102**, 103502 (2013).

[13] R. Lifshitz and M.C. Cross, *Reviews of Nonlinear Dynamics and Complexity* (Wiley-VCH Verlag GmbH & Co. KGaA, Weinheim, Germany, 2008).

[14] X. Dong, M.I. Dykman, and H.B. Chan, Nat. Commun. **9**, 3241 (2018).

[15] A. Croy, D. Midtvedt, A. Isacsson, and J.M. Kinaret, Phys. Rev. B **86**, 235435 (2012).

[16] P. Prasad, N. Arora, and A.K. Naik, Nano Lett. **19**, 5862 (2019).

[17] J.P. Mathew, R.N. Patel, A. Borah, R. Vijay, and M.M. Deshmukh, Nat. Nanotechnol. **11**, 747 (2016).

[18] A.H. Nayfeh and D.T. Mook, *Nonlinear Oscillations* (Wiley-VCH Verlag GmbH, Weinheim, Germany, 1995).

[19] R. Lifshitz and M.C. Cross, Phys. Rev. B **67**, 134302 (2003).

Supplementary Information

Tunable nonlinear damping in parametric regime

**Calculation of Effective Duffing Non-linearity ($\alpha_{eff}$):**

Figure S1 describes the MoS$_2$ resonator clamed at the edge of a circular trench. An electrostatic potential ($V_g^{dc}$) applied at gate bends the membrane and tune the frequency. The applied strain using the electrostatic force also modifies the non-linear coefficients. The gate voltage can be used as a knob to control the non-linear coefficient which is discussed the following section.

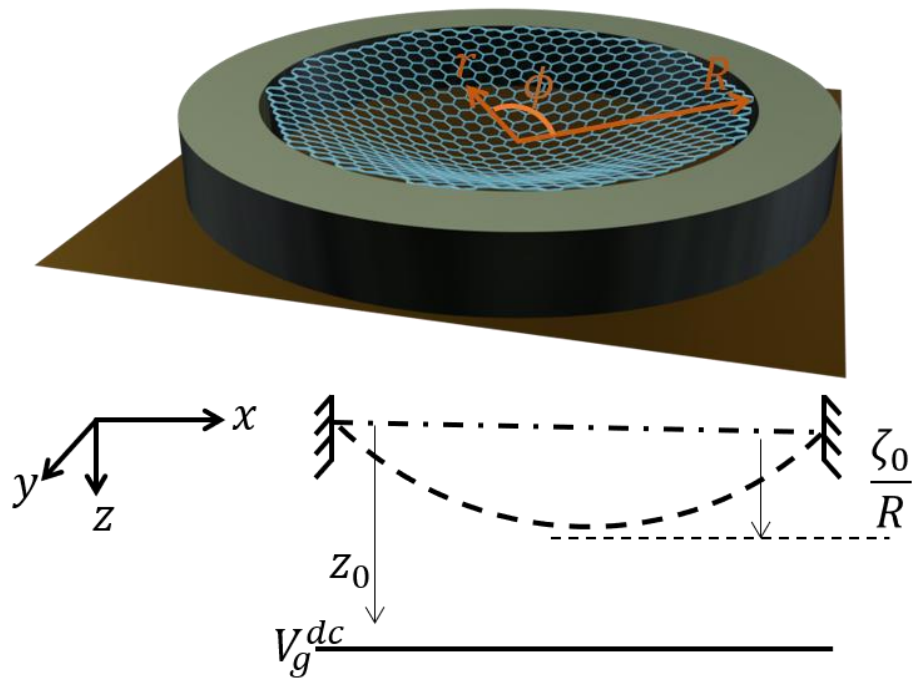

*Figure S1: MoS$_2$ NEMS: A DC voltage $V_g^{dc}$ is applied at the gate to deform the membrane. Distance between the trench and the gate is $z_0$.*

We consider continuum model of the drum resonator[1]:

$$\partial_\tau^2 \zeta_\alpha + \Lambda_\alpha \zeta_\alpha + \sum_{\beta=1}^{\infty}\sum_{\gamma\geq\beta}^{\infty} Q^\alpha_{\beta\gamma}\zeta_\beta\zeta_\gamma + \sum_{\beta=1}^{\infty}\sum_{\gamma\geq\beta}^{\infty}\sum_{\eta\geq\gamma}^{\infty} C^\alpha_{\beta\gamma\eta}\zeta_\beta\zeta_\gamma\zeta_\eta = f_\alpha(\tau) \quad (S1.1)$$

Where, $\zeta$ are the coordinates, $Q^\alpha_{\beta\gamma}$ and $C^\alpha_{\beta\gamma\eta}$ are the coupling constants, $f_\alpha(\tau)$ is the applied force and $\Lambda_\alpha$ are the eigen frequencies squared.

The maximum scaled deflection ($\zeta_0$) at the center of the resonator is given by:

$$\zeta_0 = b\left(\frac{2}{3\theta}\right)^{\frac{1}{3}} - \left(\frac{\theta}{18c^3}\right)^{\frac{1}{3}} \quad (S1.2)$$

$$\theta = \sqrt{3(4(b\ c)^3 + 27a^2c^4)^{1/2} - 9a\ c^2} \quad (S1.3)$$

Where $a = -\dfrac{R\,\epsilon_0\, V_g^{dc^2}}{4Yh\, d\, z_0^2}$, $b = \dfrac{2T_0}{Yh} - \dfrac{R^2\epsilon_0 V_g^{dc^2}}{3Y\, h\, z_0^3}$ and $c = \dfrac{7-\nu}{6(1-\nu)}$

Which translates the Stress ($T$) in the membrane as:

$$T = T_0\left(1 + \frac{z^2}{4}\right) \quad (S1.4)$$

Where $z$ is scaled static center deflection and given by:

$$z = \zeta_0 \left(\frac{Y\,h}{T_0}\right)^{\frac{1}{2}} \sqrt{\left(\frac{3-\nu}{1-\nu}\right)} \quad (S1.5)$$

The above information is used to find the non-linear coefficients.

$$\alpha_3 = C_1 \frac{T}{R^4 \rho_0} \quad (S1.6)$$

$$\alpha_2 = Q_1 \frac{T}{R^3 \rho_0} \quad (S1.7)$$

Where $C_1$ and $Q_1$ are the coupling constant determined as follows:

$$C_1 = \frac{Y\,h}{T}(3.92 + 3.68\,\delta\nu) \quad (S1.8)$$

And

$$Q_1 = \zeta_0 \frac{Y h}{T}(11.7 + 11.3\, \delta v) \qquad (S1.9)$$

After the determination of the $\alpha_2$ and $\alpha_3$, $\alpha_{eff}$ can be defined as:[2]

$$\alpha_{eff} = \alpha_3 - \frac{10}{9}\left(\frac{\alpha_2}{\omega_0}\right)^2 \qquad (S1.10)$$

We used the following parameters for the calculation of $\alpha_{eff}$

| Parameters | Value |
|---|---|
| Young's modulus | $0.33 \times 10^{12}$ Pa |
| Mass density of $MoS_2$ ($\rho$) | $5.06 \times 10^3$ kg. m$^{-3}$ |
| Poisson ratio ($\nu$) | 0.27 |

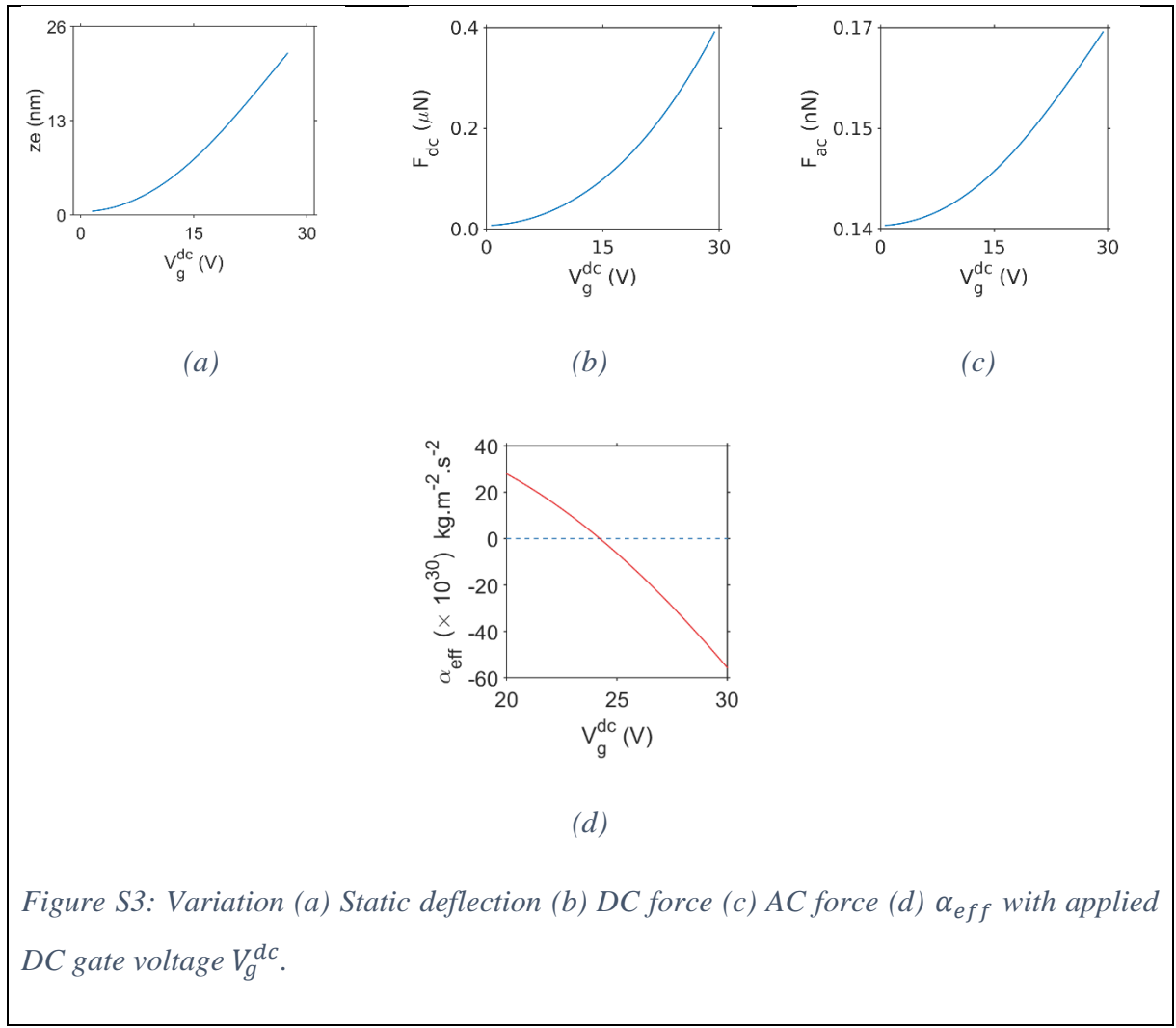

*Figure S3: Variation (a) Static deflection (b) DC force (c) AC force (d) $\alpha_{eff}$ with applied DC gate voltage $V_g^{dc}$.*

**Calculation of non-linear damping coefficient:**

We consider the following equation to describe the parametric drive[3]:

$$m\ddot{z} + \gamma \dot{z} + m\omega_0^2 z + \eta \dot{z} z^2 + \alpha_{eff} z^3 = F(\omega t) + F_p(2\omega t) z \qquad (S2.1)$$

Equation S2.1 describes nano mechanical Duffing resonator with linear and non-linear damping driven by a direct force $F$ and a parametric force $F_p$ acting at $\omega$ and $2\omega$ respectively. The resonator has mass $m$ and resonance frequency $\omega_0$. $\alpha_{eff}$, $\gamma$, and $\eta$ are Duffing, linear and non-linear damping coefficients respectively. Displacement amplitude ($z$) is a time ($t$) varying function and dot represents derivative with respect to time.

Coefficients of non-linear damping ($\eta$) can be derived from equation S1 and written as follows:

$$\eta = \frac{2 F_p Q - 4}{z^2} \qquad (S2.2)$$

Where the vibration amplitude ($z$) is scaled as follows:

$$\left(\frac{m\omega_0^2}{\tilde{\alpha}}\right)^{\frac{1}{2}} z = \tilde{z} \qquad (S2.3)$$

$$\frac{\tilde{\eta}\omega_0}{\alpha} = \frac{2 F_p Q - 4}{\tilde{z}^2 \frac{\alpha}{m\omega_0^2}} \qquad (S2.4)$$

$$\tilde{\eta} = \frac{(2 F_p Q - 4) m \omega_0}{\tilde{z}^2} \qquad (S2.5)$$

Where:

$$F_p = \frac{2 A \epsilon_0 V_g^{dc} V_p^{ac}}{\tilde{z}_0^3} \qquad (S2.6)$$

Non-linear damping constant can be estimated using equation S2.5.

**References:**

[1] A.M. Eriksson, D. Midtvedt, A. Croy, and A. Isacsson, Nanotechnology **24**, 395702 (2013).

[2] A.H. Nayfeh and D.T. Mook, *Nonlinear Oscillations* (Wiley-VCH Verlag GmbH, Weinheim, Germany, 1995).

[3] R. Lifshitz and M.C. Cross, *Reviews of Nonlinear Dynamics and Complexity* (Wiley-VCH Verlag GmbH & Co. KGaA, Weinheim, Germany, 2008).